
\font\rm=cmr12
\font\bf=cmbx12
\font\it=cmti12
\rm
\baselineskip 24pt
\hsize 6in
\hoffset=.25in
\vsize 8.5in
\
\vskip .5in
\centerline{\bf q$^2$-Dependence of Meson Mixing in Few-Body
Charge Symmetry Breaking:}
\centerline{\bf $\pi^o-\eta$ Mixing to One Loop in
Chiral Perturbation Theory}
\vskip .5in
\centerline{Kim Maltman}
\vskip .25in
\centerline{Department of Mathematics and Statistics, York
University}
\centerline{4700 Keele St., North York, Ontario  CANADA  M3J 1P3}
\centerline{and}
\centerline{PDO, Los Alamos National Lab, Los Alamos, N.M. USA
87545}
\vskip 1in
\centerline{ABSTRACT}
\vskip .25in
\noindent
It is pointed out that the meson mixing matrix elements usually
considered responsible for the bulk of the observed few-body
charge symmetry breaking are naturally $q^2$-dependent in
QCD.  For $\pi^o-\eta$ mixing,
using the usual representation of the pseudoscalar fields,
the leading $q^2$ dependence
can be explicitly calculated using chiral perturbation theory
to one loop, the result being a significant decrease in the
magnitude of the matrix element in going from timelike to
spacelike values of $q^2$.  Since it is the latter range of
$q^2$ which is relevant to NN scattering and the few-body
bound state, this result calls into serious question the
standard treatment of few-body charge symmetry breaking
contributions associated with $\pi^o-\eta$ and $\rho -\omega$
mixing.
\vfill\eject

In the standard model, charge symmetry (CS) is broken not
only by electromagnetism (EM) but also, through the inequality
of up and down quark masses, by the strong interactions.
This strong breaking leads to important non-Coulombic
contributions to CS violating observables such as the difference
of low energy $nn$ and $pp$ scattering amplitudes and the $A=3$
binding energy difference, $\Delta BE$.$^{1-3}$  It is generally
believed that these non-Coulombic contributions can be accounted
for in the meson exchange picture by isoscalar-isovector mixing
(predominantly $\rho -\omega$, but also $\pi^o-\eta$) in the
intermediate meson propagators of one-boson exchange
graphs.$^{4-6}$  In the analysis of Refs 4-6, the $\rho -\omega$
matrix element is obtained from experimental data on
$e^+e^-\rightarrow \pi^+\pi^-$ in the region of the
$\rho -\omega$ interference shoulder, while the $\pi^o-\eta$
matrix element is obtained by a pole model analysis of
$\eta$, $\eta^\prime$ decays.  The latter analysis (which
has some additional shortcomings, to be discussed below)
is feasible only because the mixing
matrix elements are assumed to be $q^2$-independent.
This assumption is
also implicit in the treatment of the $A=3$ bound state and
NN scattering, where timelike matrix elements, extracted
as above, are used unchanged in the spacelike region.
{}From the perspective of QCD, however, this assumption is
clearly incorrect.  Indeed, the basic hypothesis underlying
the meson exchange picture is that, at low energies, QCD
reduces to an effective hadronic theory involving mesons,
nucleons, deltas (etc.).  We know, however, that, as an
effective low energy theory, such a hadronic theory must
incorporate all possible terms, involving the low energy
composite fields, which are not explicity forbidden by
the symmetries of the underlying theory (in this case, QCD).
In particular, there will be terms higher order in derivatives,
involving also the quark mass matrix, which will produce
a $q^2$-dependence to all isoscalar-isovector mixing matrix
elements.  How important such effects will be for a given
CS-breaking (CSB) observable cannot be determined without either
generating the effective theory from QCD or constraining
the strengths of the relevant terms in the effective
Lagrangian from experiment.  Some information, however,
already suggests that these effects are unlikely to be
negligible.  First, we know that, in the case of effective
chiral Lagrangians, higher derivative terms typically
occur scaled by a dimensionful factor $\Lambda_\chi\sim 4\pi
 f_\pi\sim 1 GeV$.$^{7-10}$  If this is true in general,
one would expect significant changes in going from,
eg., experimental data for $\rho -\omega$ at $q^2\sim m_\rho^2$,
to $q^2 < 0$.  (We will consider the case of $\pi^o-\eta$
mixing, using the chiral Lagrangian, explicitly below and
see that this expectation is borne out.)  Second, a
recent model calculation$^{11}$, in which the $q^2$-dependence
of $\rho -\omega$ mixing arising from an intermediate (free)
quark loop through the $u-d$ constituent quark mass
difference is evaluated using a monopole {\it ansatz} for
the meson-quark-antiquark vertex function, finds a very
strong $q^2$ dependence of the $\rho -\omega$ element
of the inverse vector meson propagator matrix.  Such
a $q^2$ dependence would translate into a significant
$q^2$ dependence of the phenomenological $\rho -\omega$
mixing matrix element (this aspect of the calculation is
not treated in Ref 11--see Ref 12 for a detailed discussion).
This result is qualitatively
compatible with the discussion above and, despite possible
reservations about its potential model-dependence, makes
clear the dangers inherent in the neglect of $q^2$-dependence
of the usual treatment.

One would, of course, ideally, like to reduce the model
dependence of the $\rho -\omega$ result, since it is
$\rho -\omega$ mixing which produces the dominant non-Coulombic
CSB contribution to few-body observables (in the standard
treatment).  Unfortunately, one is hampered, in this regard,
by a lack of information about higher order terms in the
(putative) effective low-energy Lagrangian involving the
vector meson fields.  If we turn our attention
to the next-to-leading contribution, associated with $\pi^o-\eta$
mixing, however, we are in much better shape, since the
effective low-energy Lagrangian relevant to the pseudoscalar
sector is well-known, to order $q^4$, as a result of the work
of Gasser and Leutwyler$^{8-10}$.  By computing $\pi^o-\eta$
mixing to one-loop in this framework, one obtains the leading
$q^2$-dependence in a straightforward, and reliable, fashion.
Since the effective Lagrangian is not constrained beyond order
$q^4$, one cannot determine the higher order $q^2$-dependent
contributions, which restriction limits the validity of
the results to the range $\vert q^2\vert < O(m_\eta^2)$, by
the usual power counting arguments of the low-energy
expansion.

Let us consider, therefore, $\pi^o-\eta$ mixing.  Recall that
the relevant terms in the effective chiral Lagrangian,
to order $q^4$, are given by$^8$
$$\eqalignno{
L_{eff}\ =\ &{1\over 4}f^2 Tr(\partial_\mu\Sigma\partial^\mu
\Sigma^\dagger )
+{1\over 2}f^2 Tr[\mu M(\Sigma +\Sigma^\dagger )]
+L_1 \bigl[ Tr(\partial_\mu\Sigma\partial^\mu\Sigma^\dagger )
\bigr]^2
\cr
&+ L_2 Tr(\partial_\mu\Sigma\partial_\nu\Sigma^\dagger )
Tr(\partial^\mu\Sigma\partial^\nu\Sigma^\dagger )
+ L_3 Tr(\partial_\mu\Sigma^\dagger\partial^\mu\Sigma
\partial_\nu\Sigma^\dagger\partial^\nu\Sigma )\cr
&+ L_4 Tr(\partial_\mu\Sigma\partial^\mu\Sigma^\dagger )
Tr[2\mu M(\Sigma +\Sigma^\dagger )]
+ L_5 Tr\bigl[ 2\mu (M\Sigma +\Sigma^\dagger M)
\partial_\mu\Sigma^\dagger \partial^\mu\Sigma \bigr]\cr
&+ L_6 \bigl[ Tr[2\mu M(\Sigma +\Sigma^\dagger )]\bigr]^2
+ L_7 \bigl[ Tr[2\mu M(\Sigma -\Sigma^\dagger )]\bigr]^2\cr
&+L_8 Tr[4\mu^2(M\Sigma M\Sigma +M\Sigma^\dagger M\Sigma^\dagger )]
&(1)\cr}$$
where $\mu$ is a mass scale
related to the value of the quark condensate$^8$,
$$\Sigma = exp(i\vec \lambda . \vec \pi /f) ,\eqno(2)$$
with $\vec\lambda$ the usual $SU(3)$ Gell-Mann matrices,
$\vec\pi$ the pseudoscalar octet fields and $f$ a dimensionful
constant, equal to $f_\pi$ in leading order.  In Eq (1),
$M$ is the current quark mass matrix and we have set the
external fields to zero.  The
(scale-dependent) coeffieients $L_1,\cdots , L_8$ are as
defined in Ref 8.  In what follows we ignore EM contributions
to $\pi^o-\eta$ mixing, since these are known to vanish in the
chiral limit$^{13}$ and estimates for the effect of the
departure from the chiral limit, based on the size of the
relevant EM chiral logarithms$^{14}$, show them to be
negligible relative to the strong contributions resulting
from $L_{eff}$ in Eq (1).  From Eq (1) one may compute the
inverse $\pi_3-\pi_8$ propagator matrix to one-loop as a
function of $q^2$ (where $q$ is the external four-momentum
and $\pi_3$, $\pi_8$ are the unmixed octet fields that would
be identical to the
$\pi^o$, $\eta$ fields, respectively, in the limit of
exact isospin symmetry).  Diagonalizing this matrix one
obtains the $q^2$-dependent mixing angle for the physical
propagating modes
\vfill\eject
$$\eqalignno{\theta (q^2)\ =\ &{\sqrt3 (m_d-m_u) \over 4(m_s-
\hat m )}\Biggl[ 1-{32\mu (m_s-\hat m )\over f^2}
(3L_7^r+L_8^r)
+{(-3\ell_\pi +\ell_\eta
+2\ell_K )\over 32\pi^2 f^2}\cr
&+\Bigl( {q^2+m_\eta^2 \over 32\pi^2 f^2}\Bigr)
\Bigl( 1+{m_\pi^2 \over (m_K^2-m_\pi^2)}\ell n(m_\pi^2/m_K^2)
\Bigr) \Biggr] &(3)\cr}$$
where $m_u, m_d, m_s$ are the $u,d,s$ current quark masses,
$\hat m = (m_u+m_d)/2$, $L_i^r$ are the (scale-dependent)
renormalized low-energy constants of Ref 8 and
$\ell_P = m_P^2 \ell n(m_P^2/\mu_o^2)$, with $P$ any of
the pseudoscalar mesons and $\mu_o$ the renormalization
scale.  The $q^2$-dependence of $\theta$ is unavoidable at
the one-loop level.  Following Ref 8 one may rewrite
Eq (3) using one-loop results for other physical
observables, obtaining
$$\eqalignno{\theta (q^2)\ =\ &{\sqrt3 (m_{K^o}^2-m_{K^+}^2)_{QCD}
\over (m_K^2-m_\pi^2)}\Biggl[ 1+\Delta_{GMO}\cr
&+{1 \over 16\pi^2 f^2}\left( {m_\eta^2 \over (m_\eta^2-m_\pi^2)}
\right)
\left( 3m_\eta^2\ell n(m_K^2/m_\eta^2)+m_\pi^2\ell n(m_K^2/
m_\pi^2)\right) \cr
&+\left( {q^2+m_\eta^2 \over 32\pi^2 f^2}\right) \left( 1+
{m_\pi^2 \over (m_K^2-m_\pi^2)}\ell n(m_\pi^2/m_K^2)\right)
\Biggr]&(4)\cr}$$
where $\Delta_{GMO}$ is the (electromagnetically corrected)
Gell-Mann-Okubo discrepancy
$$\Delta_{GMO}=(4m_K^2-m_\pi^2-3m_\eta^2)/(m_\eta^2-m_\pi^2).
\eqno(5)$$
In Eq (4), $(\Delta m_K^2)_{QCD}$ is the contribution to the
$K^o-K^+$ mass-squared splitting resulting from $m_u\not= m_d$.
This can be extracted from the physical splitting if one knows
the corresponding EM contribution.  The latter may be considered
to be known if we accept Dashen's theorem$^{13}$ for the EM
self-energies,
$$(\Delta m_K^2)_{EM}=(\Delta m_\pi^2)_{EM}\eqno(6)$$
since the quark-mass-difference contribution to the $\pi$
splitting is $O[(m_d-m_u)^2]$ and known to be small$^{15}$.
Dashen's theorem is, however, valid only to lowest order in
the EM chiral expansion.  Constraints on the size of potential
violations are discussed in Ref 16.  The resulting
uncertainties in the size of $(\Delta m_K^2)_{EM}$ lead to
a corresponding uncertainty of order 20\% in the overall scale
of the result (4).
Apart from this overall scale uncertainty, the result (4) should
provide an accurate representation of $\theta (q^2)$, subject
to the restrictions of the validity of the low energy expansion,
i.e. for values of $\vert q^2\vert < O(m_\eta^2)$.

Before proceeding, we must point out one important feature
of the calculation which bears crucially on the proper
interpretation of its results.  Recall that $L_{eff}$ is only
one of a family of possible effective Lagrangians, all
others being related to that in Eqn (1) by redefinitions of
the octet fields which leave the one particle singularities,
and hence the S-matrix elements, of the theory unchanged$^{17)}$.
The propagator matrix, and hence the mixing angle,
$\theta (q^2)$, of course, depends on the particular
definition of the octet fields employed:  as such it
is not a physical quantity.  Only when combined with
a full low-energy Lagrangian including the nucleon
fields could one obtain results for physical processes
(such as, eg., NN scattering) which were independent of
the particular choice of the meson fields.  In the context
of few-body meson-exchange model calculations we are not,
at present, able to carry out this program and, in fact,
because of the insertion of phenomenological form
factors for the meson-nucleon vertices, it is actually
ambiguous as to exactly what definition of the meson
fields is being used (in the sense above).  The relevant
question, in this setting, is then whether or not the
$q^2$ dependence of the mixing (which must be present,
on general grounds, for any choice of fields) is typically
negligible or not.  If we use the definition of the fields
which leads to $L_{eff}$ of the form given in Eqn (1),
we find that $q^2$ dependence of the mixing is not negligible
in few-body systems, and hence that the assumption of
$q^2$ independence of the standard treatments is not a
reasonable one to make.  Unless, however, one formulates
the entire scattering (or bound state) problem under
consideration in a well-defined effective field theoretic
manner, where one can determine that the same choice for
the meson fields is being used for the meson-nucleon
vertices as in the purely mesonic sector, one is not
guaranteed that simply plugging the result of Eqn (7)
into an existing few-body code will lead to a reliable
estimate of the CSB associated with $\pi^o -\eta$ mixing.
This statement is, of course, equally true of the usual
treatment, with, however, the added caveat that the usual treatment
suffers from the absence of a
demonstration that there even  exists a choice
of the meson fields for which the mixing is $q^2$-independent.
The result of Eqn (7) is, of course, perfectly reliable
for the choice of meson fields implied by Eqn (1); the
problem is that possible additional $q^2$ dependent CSB
contributions
associated with the meson-nucleon couplings cannot be
{\it consistently} evaluated owing to the phenomenological
treatment of the meson-nucleon vertices.
For this reason, quantities such as the $\pi^o-\eta$ and
$\rho -\omega$ matrix elements used in conventional
few-body CSB calculations are actually ambiguous beyond
leading order.  This means that the appropriate (and
most conservative) interpretation of the present calculation
is that the difference between the results of using
the $q^2$-dependent mixing of Eqn (7) and the usual
$q^2$-independent mixing represents a lower bound for
the {\it uncontrollable} theoretical uncertainties
associated with the phenomenological modelling of
the input meson-nucleon vertices.  The comments which
follow are to viewed in this light.

{}From the results above we obtain, for the choice of
fields implied by Eqn (1), two pieces of information:
first, the size of the $\pi^o-\eta$ mixing and, second, its
$q^2$ dependence.

Let us first consider the magnitude.  The result (4) leads to
a phenomenological mixing matrix element given by
$$m_{38}^2(q^2) =  -(m_\eta^2-m_\pi^2)\theta (q^2) .\eqno(7)$$
As mentioned above, there is an overall uncertainty in the
result (7) as a consequence of the unknown size of possible
violations of Dashen's theorem.  Since this uncertainty
also enters the determination of $\Delta_{GMO}$ the effect
is not quite linear in $(\Delta m_K^2)_{QCD}$, but it is
very nearly so.  EM and strong contributions to $\Delta m_K^2$
have the opposite sign so that EM contributions larger than
those corresponding to Dashen's theorem lead to larger
values of $(\Delta m_K^2)_{QCD}$.  Let us write
$$m_{38}^2=-[a_o +(q^2/m_\eta^2)a_1].\eqno(8)$$
The quantities $a_o, a_1$
are listed in Table 1 for a range of values of
$(\Delta m_K)_{EM}$ between $1.3 MeV$ (arising from
Dashen's theorem) and $2.8 MeV$ (which results from attempting
to saturate the Cottingham formula for the kaon system
with $K$ and $K^*$ intermediate states$^{18}$).

The values of Table 1 should be considered to supercede
those of earlier work on the subject for the following
reasons.  The $\pi^o-\eta$ matrix element of Ref 6c was
extracted by an analysis of $\eta$, $\eta^\prime$ decays
in which these decays are taken to be mediated by $\eta$,
$\eta^\prime$ and $\pi$ poles, intermediate $\pi^o\eta$,
$\pi^o\eta^\prime$ vertices being taken to be $q^2$-independent.
There are two problems with this analysis.  First, as we
have seen, the $\pi^o-\eta$ vertex (and hence, presumably,
also the $\pi^o-\eta^\prime$ vertex) is not $q^2$-independent.
Second, as one discovers from Ref 10, which treats
$\eta\rightarrow 3\pi$ to one loop in chiral perturbation
theory, a large portion of the $55\%$ increase of the
$\eta\rightarrow 3\pi$ amplitude in going from tree level
to one-loop level is attributable to enhancements associated
with final state interactions produced by
the presence of $s$-wave $I=0$ $\pi\pi$ pairs.  As an
$s$-wave $\pi\pi$ effect this
enhancement can clearly not be represented by an intermediate
pseudoscalar pole (the effect of the $\eta^\prime$ in
$L_{eff}$ is contained entirely in the low-energy
constant $L_7$\ $^{8)}$);
it has, therefore, not been included in
the pole model analysis.  Its inclusion would presumably lower the
extracted values of the $\pi^o-\eta$ and $\pi^o-\eta^\prime$
matrix elements, even if one could treat them as $q^2$-independent.
There is, however, no need to perform a modified pole model
analysis to extract the $\pi^o-\eta$ matrix element, since
we can obtain it directly from chiral perturbation theory, at
least for the region of $q^2$ for which the low energy
expansion is valid.\footnote*{It should be noted that
the analysis of Refs 6a,6b which uses the quark model,
together with certain results valid in the chiral limit,
to evaluate, eg., the $\pi^o-\eta$ matrix element, contains
non-vanishing EM contributions which are incorrect.  The
EM contribution to $\pi^o-\eta$ mixing actually vanishes
in the chiral limit$^{13}$.  The error results from neglecting
a class of graphs (\lq\lq EM penguins of the
second kind"$^{19}$) which are required to properly implement all
of the EM chiral constraints of Ref 13, in the chiral limit.
The numerical consequence of this error are, however,
not particularly significant.}

Next we turn to the $q^2$-dependence of $m_{38}^2$.  If
we consider the change in the matrix element in going
from $q^2=m_\eta^2$ to $q^2=-m_\eta^2$ (very nearly equal
to the range over which the $\rho -\omega$ matrix element
is extrapolated to get from the $e^+e^-$ experimental data
to the scattering region) we find
that there is a decrease of $17\%$.  One of the advantages
of a model calculation would be that the model would
provide also the higher order $q^2$ dependence;  here
we have access only to the leading (linear-in-$q^2$) dependence,
though this dependence has (apart from the uncertainty
in the overall scale, a feature subject to future improvement)
the advantage of theoretical reliability, in the sense that
the structure of the low-energy expansion of chiral
perturbation theory is directly related to QCD and its
convergence to one-loop has been tested for a large number
of processes$^{19}$.  We see that the $q^2$ dependence
of the mixing in the meson propagator
is
certainly not negligible.  While one would like to quote
numbers on how much this affects, eg., the uncertainty
in the $\pi^o-\eta$
contribution to the $A=3$ binding energy difference
(in the conservative sense described above), this
is not possible because the binding energy difference is,
in fact, sensitive to spacelike $q^2$ values considerably
outside the range of validity of the low-energy expansion$^3$.
In order to make further progress one should, therefore,
use the chiral results embodied in Eqs (7) and (8) to constrain
model calculations.  If an extension of the calculation of
Ref 11 to the pseudoscalar sector were to reproduce the
$q^2$-dependence of the $\pi^o-\eta$ matrix element, this
would considerably enhance our confidence in what it has to
say about the running of the $\rho -\omega$ matrix element.
\vfill\eject
\
\vskip 1in
\centerline{\bf ACKNOWLEDGEMENTS}
\vskip .25in
\noindent
This work was supported by the Natural Sciences and
Engineering Council of Canada and the U.S. Department
of Energy.  Much of the work reported here was performed
at the Institute for Nuclear Theory, in conjunction with
the workshop \lq\lq Electromagnetic Interactions and the
Few-Nucleon Systems\rq\rq .  The author would like to
thank the Institute for its support and hospitality and
to acknowledge useful conversations with Jim Friar, Sid
Coon, Terry Goldman and Jerry Stephenson.
\vfill\eject
\centerline{\bf REFERENCES}
\vskip .4in
\noindent
1.  G.A. Miller, B.M.K. Nefkens and I. Slaus, Phys. Rep.
{\bf 194} (1990) 1.

\noindent
2.  J.L. Friar, Nucl. Phys. {\bf A156} (1970) 43; M. Fabre
de la Ripelle, Fizika {\bf 4} (1972) 1; J.L. Friar and
B.F. Gibson, Phys. Rev. {\bf C18} (1978) 908; J.L. Friar,
B.F. Gibson and G.L. Payne, Phys. Rev. {\bf C35} (1987) 1502.

\noindent
3.  R.A. Brandenburg, S.A. Coon and P.U. Sauer, Nucl. Phys.
{\bf A294} (1978) 1752.

\noindent
4.  P.C. McNamee, M.D. Scadron and S.A. Coon, Nucl. Phys.
{\bf A249} (1975) 483.

\noindent
5.  S.A. Coon and R.C. Barrett, Phys. Rev. {\bf C36} (1987) 2189.

\noindent
6.  a)  H.F. Jones and M.D. Scadron, Nucl. Phys. {\bf B155}
(1979); b) M.D. Scadron, Phys. Rev. {\bf D29} (1984) 2076;
c) S.A. Coon, B.H.J. McKellar and M.D. Scadron, Phys. Rev.
{\bf D34} (1986) 2784.

\noindent
7. A. Manohar and H. Georgi, Nucl. Phys. {\bf B234} (1984) 189.

\noindent
8.  J. Gasser and H. Leutwyler, Nucl. Phys. {\bf B250} (1985) 465.

\noindent
9.  J. Gasser and H. Leutwyler, Nucl. Phys. {\bf B250} (1985) 517.

\noindent
10. J. Gasser and H. Leutwyler, Nucl. Phys. {\bf B250} (1985) 539.

\noindent
11. T. Goldman, J.A. Henderson and A.W. Thomas, Los Alamos
preprint LA-UR-91-2010, 1991.

\noindent
12. K. Maltman and T. Goldman, ``A Quark-Loop Model for
the Off-Shell Dependence of $\rho -\omega$ Mixing'',
in preparation.

\noindent
13. R. Dashen, Phys. Rev. {\bf 183} (1969) 1245.

\noindent
14. K. Maltman, Phys. Rev. {\bf D44} (1991) 751.

\noindent
15. J. Gasser and H. Leutwyler, Phys. Rep. {\bf 87C} (1982) 77.

\noindent
16. K. Maltman and D. Kotchan, Mod. Phys. Lett. {\bf A5} (1990)
2457.

\noindent
17. S. Coleman, J. Wess and B. Zumino, Phys. Rev. {\bf 177}
(1969) 2239.

\noindent
18. R.H. Socolow, Phys. Rev. {\bf 137} (1965) 1221.

\noindent
19. K. Maltman, G.J. Stephenson Jr. and T. Goldman, Nucl. Phys.
{\bf A539} (1991) 539; G.J. Stephenson Jr., K. Maltman and
T. Goldman, Phys. Rev. {\bf D43} (1991) 860.

\noindent
20. J.F. Donoghue and B.R. Holstein, Phys. Rev. {\bf D40}
(1989) 2378 and Phys. Rev. {\bf D40} (1989) 3700.  See
also Ref 8.
\vfill\eject
\
\vskip 2in
\noindent
Table 1.  The constants $a_o$, $a_1$ as a function
of $(\Delta m_K)_{EM}$\ $^*$
\vskip .5in
\noindent
\hrule
\settabs\+\qquad\qquad &$(\Delta m_K)_{EM}$ (MeV)\qquad\qquad
&$a_o$ (GeV$^2$)\qquad\qquad&$a_1$ (GeV$^2$)\qquad&\cr
\vskip .15in
\noindent
\+ &$(\Delta m_K)_{EM}$ (MeV)
&$a_o$ (GeV$^2$)&$a_1$ (GeV$^2$)&\cr
\vskip .15in
\noindent
\hrule
\vskip .2in
\noindent
\+&\ \ \ 1.3&.00335&.00028&\cr
\+&\ \ \ 1.6&.00354&.00030&\cr
\+&\ \ \ 1.9&.00372&.00032&\cr
\+&\ \ \ 2.2&.00390&.00033&\cr
\+&\ \ \ 2.5&.00408&.00035&\cr
\+&\ \ \ 2.8&.00426&.00036&\cr
\vskip .1in
\noindent
\hrule
\vskip .2in
\noindent
$^*$ All notation as in the text.  $a_o$ and $a_1$ are as
defined in Eq (8).
\vfill\eject
\bye